\documentstyle[12pt]{article}

\author{Yu. M. Zinoviev
        \thanks{E-mail address: ZINOVIEV@MX.IHEP.SU} \\
        {\it Institute for High Energy Physics} \\
        {\it Protvino, Moscow Region, 142284, Russia}}
\title{Spontaneous symmetry breaking \\
       in $N=2$ supergravity}
\date{}

\begin{document}

\maketitle

\begin{flushright}
\vspace*{-2.7in}
IHEP 86-126 \\
 May 1986 \\
\end{flushright}
\vspace{2in}

\begin{abstract}
 A model describing the $N=2$ supergravity interaction with vector and
linear multiplets is constructed. It admits the introduction of the
spontaneous breaking of supersymmetry with arbitrary scales, one of
which may be equal to zero, which corresponds to partial super-Higgs
effect ($N=2 \rightarrow N=1$). A cosmological term is automatically
equal to zero.
\end{abstract}

  Up to now supersymmetric phenomenology has, as a rule, been
constructed on the basis of the $N=1$ supergravity \cite{r1}. A
possibility to construct realistic models, based on extended
supergravities is first of all connected with the solution of the
problem on spontaneous supersymmetry breaking. In this, the most
attractive scenario is the one, where the extended supergravity $N > 1$
would be broken up to $N=1$.

  As is shown in \cite{r2} the $N=2$ supergravity with vector multiplets
admits spontaneous symmetry breaking without a cosmological term (the
case of "flat" potentials) however in such models the $N=2$
supersymmetry is completely broken, in this gravitini being mass
degenerate.

  This paper is devoted to the study of possibilities of spontaneous
symmetry breaking in the $N=2$ supergravity with two different scales,
including the case of partial super-Higgs effect, when the $N=1$
supersymmetry remains unbroken (and the corresponding gravitino being
massless). It has been pointed out in \cite{r3} that partial breaking of
$N=2 \rightarrow N=1$ would lead to the appearence of a massive
$(3/2,1,1,1/2)$ supermultiplet, which means that the problem reduces to
a possibility to introduce an interaction of such a multiplet with the
$N=1$ supergravity. Since the interaction of supergravity with matter is
naturally constructed in terms of massless multiplets, one should use
such a description of the massive supermultiplet which admits a
nonsingular massless limit. Such a formalism can easily be constructed
if one uses a gauge invariant description of spin $3/2$ and $1$
\cite{r4}, which has been done in Section 1. The results on analysing
possibilities for such a multiplet to appear in the $N=2$ supergravity
are given in the same Section. In Section 2 one can find a complete
non-linear interaction Lagrangian for a minimal set of multiplets,
satisfyng the requirements of Section 1. In Section 3 it is shown in what
way a supersymmetry breaking with two arbitrary scales may be introduced
into such a system, $N=2 \rightarrow N=1$ breaking being possible as a
particular case. In the given minimal model the scalar field potential
does not arise at all, consequently the cosmological term is
automatically equal to zero.

\section{}

  Within the massless limit a massive particle with spin $3/2$ should
decay into massless particles with spin $3/2$ and $1/2$ in the same way
as a massive spin particle with spin $1$ --- into massless ones with
spin $1$ and $0$. Simple counting of the degrees of freedom shows that a
massive $(3/2,1,1,1/2)$ supermultiplet in the massless limit should
reduce to three massless supermultiplets $(3/2,1)$, $(1,1/2)$,
$(1/2,0,0)$, respectively.

   The Lagrangian describig a free supermultiplet, has the form
\begin{eqnarray}
 L_o &=& \frac{i}2 \varepsilon^{\mu\nu\alpha\beta} \bar{\Psi}_{\mu}
\gamma_5 \gamma_\nu \partial_\alpha  \Psi_\beta - \frac14 A^2_{\mu\nu}
 - \frac14 B^2_{\mu\nu} + \nonumber \\
  && + \frac{i}2 \bar{\Omega} \hat{\partial} \Omega + \frac{i}2 \bar{\chi}
\hat{\partial} \chi + \frac12 (\partial_\mu \varphi)^2 + \frac12
(\partial_\mu \pi)^2 + \nonumber \\
  && + \frac{m^2}2 (A^2_\mu + B^2_\mu) - m(A^\mu \partial_\mu \varphi +
B^\mu \partial_\mu \pi) +  \nonumber \\
  && + \left[ \frac12 \bar{\Psi}_\mu \sigma^{\mu\nu} \Psi_\nu -
\frac{i}{\sqrt{2}} (\bar{\Psi} \gamma) \Omega + i (\bar{\Psi} \gamma )
\chi - \sqrt{2} \bar{\Omega} \chi + \frac12 \bar{\chi} \chi \right]
\end{eqnarray}
where spinors $\psi_\mu$, $\Omega$, $\chi$ are majorana, $A_{\mu\nu} =
\partial_\mu A_\nu - \partial_\nu A_\mu$ etc. This Lagrangian is
invariant under the following global supertransformations:
\begin{eqnarray}
 \delta \Psi_\mu &=& - \frac{i}4 \sigma (A - \gamma_5 B) \gamma_\mu \eta
- \partial_\mu (\varphi + \gamma_5 \pi) \eta + m(A_\mu + \gamma_5 B_\mu)
\eta  \nonumber \\
 \delta A_\mu &=& (\bar{\Psi} \eta) + \frac{i}{\sqrt{2}} (\bar{\Omega}
\gamma_\mu \eta ) \qquad \delta B_\mu = (\bar{\Psi} \gamma_5 \eta )
+ \frac{i}{\sqrt{2}} (\bar{\Omega}\gamma_\mu \gamma_5 \eta ) \\
\delta \Omega  &=& - \frac1{2\sqrt{2}} \sigma (A + \gamma_5 B) \eta
\qquad \delta \chi  = -i \gamma^\mu \left[ \partial_\mu (\varphi +
\gamma_5 \pi) - m(A_\mu + \gamma_5 B_\mu) \right]  \eta \nonumber \\
 \delta \varphi  &=& (\bar{\chi} \eta) \qquad \delta \pi  = (\bar{\chi}
\gamma_5 \eta ) \nonumber
\end{eqnarray}
and the local gauge transformations:
\begin{eqnarray}
 \delta \Psi_\mu &=& (\partial_\mu + \frac{im}2 \gamma_\mu)\xi \qquad
\delta \Omega  = - \frac{m}{\sqrt{2}} \xi  \qquad \delta \chi  = m \xi
\nonumber \\
 \delta A_\mu &=& \partial_\mu \Lambda_1 \qquad \delta \varphi  = m
\Lambda_1 \\
 \delta B_\mu &=& \partial_\mu \Lambda_2 \qquad \delta \pi  = m
\Lambda_2 \nonumber
\end{eqnarray}
As can easily be seen, Lagrangian (1) and supertransformation (2) do
possess a corect massless limit.

  As has already been pointed out, a possibility for partial symmetry
breaking in the $N=2$ supergravity reduces to a possibility to switch
the interaction of such a massive multiplet with the $N=1$ supergravity
(additing, if necessary, the multiplets not containing spin 3/2
particles) in a way so that the Lagrangian would be invariant with
respect to the second (spontaneously broken) local supertransformation
corresponding to $\xi$-transformation in (3).

  As usually, in supergravity the presence of scalar fields leads to the
fact that the total interaction Lagrangian and corresponding
supertransformation laws are expressed in the form of infinite series
over powers $k\varphi_i$, where $\varphi_i$ are scalar fields, and $k$
is a gravitational interaction constant (further on, put $k=1$).
However, the first itterations of switching interaction turn out to be
sufficient to determine the minimal set of massless $N=2$
supermultiplets and the peculiarities of their interaction with the
$N=2$ supergravity, which are necessary for the appearance of a partial
super-Higgs effect.

  Such an approach in investigating the spontaneous symmetry breaking in
supergravity is constructive, still, in the case of $N>1$ it also rather
cumbersome (the $N=1$ case has been considered in \cite{r5}). Therefore
we shall restrict ourselves with presenting here only the results of
such an analisys.

  The spectrum of the states in the $N=1$ supergravity with massive
$(3/2,1,1,1/2)$ supermultiplet in the massles limit precisely
corresponds to the $N=2$ supergravity with massless vector multiplet.
However, as it is indicated in \cite{r6}, this is not sufficient. The
analisys carried out makes it clear that the minimal set of multiplets
should also contain a version of a "linear" multiplet, which includes an
$SU(2)$ doublet of spinor fields $\chi_i$, $i=1,2$ and an $SU(2)$
singlet as well as a triplet of scalar fields $\tilde{\varphi}$ and
$\pi^a$, respectively. The peculiarity of vector multiplet
interaction lies in the fact that two vector fields (the field of vector
multiplets and "graviphoton") enter the Lagrangian and
supertransformations in the combinations $(A_\mu \pm \gamma_5 B_\mu)$
(similar to (2)). A characteristic feature of the new version of the
linear multiplet is the fact that scalar fields enter the Lagrangian
through the derivatives $\partial_\mu \pi^a$.

\section{}

  Let us consider the interaction of the $N=2$ supergravty $e_{\mu r}$,
$\psi_{\mu i}$, $A_\mu$, $i=1,2$, vector $B_\mu$, $\Omega_i$, $\varphi$,
$\pi$ and linear $\chi_i$, $\tilde{\varphi}$, $\pi^a$ supermultiplets.
It is convenient to introduce the notations $\pi_i{}^j = \pi^a
(\tau^a)_i{}^j$, where $\tau^a$ are standard Pauli matrices where
imaginary units are replaced by $\gamma_5$ matrices
\begin{eqnarray}
 \tau_1 &=& \left( \begin{array}{cc} O & -1 \\ 1 & O \end{array}
\right) \qquad \tau_2 = \left( \begin{array}{cc} O & \gamma_5 \\
\gamma_5 & O \end{array} \right) \qquad \tau_3 = \left(
\begin{array}{cc} \gamma_5 & O \\ O & - \gamma_5 \end{array} \right)
\end{eqnarray}
Then, for instance, $(\tau^a)_i{}^j \gamma_\mu = - \gamma_\mu
(\tau^a)_j{}^i$, etc.

  The limitations on a possible form of the interaction formulated in
the Section above turn out to be quite restrictive nevertheless one
manages to construct such a model. The corresponding interaction
Lagrangian (up to four-fermionic terms) has the form
\begin{eqnarray}
 L &=& - \frac12 R + \frac{i}2 \varepsilon^{\mu\nu\alpha\beta}
\bar{\Psi}^i_\mu \gamma_5 \gamma_\nu D_\alpha \Psi_{\beta i} + \frac{i}2
\bar{\Omega}^i \hat{D} \Omega_i + \frac{i}2 \bar{\chi}^i \hat{D} \chi_i
- \nonumber  \\
  && - \frac14 e^{\sqrt{2}\varphi} (A^2_{\mu\nu} + B^2_{\mu\nu}) +
\frac1{2\sqrt{2}} \pi (A_{\mu\nu} \tilde{A}_{\mu\nu} + B_{\mu\nu}
\tilde{B}_{\mu\nu}) + \nonumber \\
  && + \frac12 (\partial_{\mu} \varphi)^2 + \frac12 e^{-2\sqrt{2}\varphi }
(\partial_\mu \pi)^2 + \frac12 (\partial_\mu \tilde{\varphi})^2 + \frac12
e^{2\tilde{\varphi}} (\partial_\mu \pi^a)^2 + \nonumber  \\
  && + e^{\varphi/\sqrt{2}} \{ \frac14 \varepsilon^{ij} \bar{\Psi}_{\mu
i} (A^{\mu \nu } - \gamma_5 \tilde{A}^{\mu\nu} + \gamma_5 B^{\mu\nu} +
\tilde{B}^{\mu\nu} ) \Psi_{\nu j} + \nonumber \\
  && + \frac{i}{4 \sqrt{2}} \bar{\Omega}^i \gamma^\mu \sigma (A +
\gamma_5 B) \Psi_{\mu i} - \frac18 \varepsilon^{ij} \bar{\chi}_i (A -
\gamma_5 B) \chi_j \} - \nonumber \\
  && - \frac12 \varepsilon^{ij} \bar{\Omega}_i \gamma^\mu \gamma^\nu
\left[ \partial_\nu \varphi + e^{- \sqrt{2}\varphi} \gamma_5
\partial_\nu \pi  \right] \Psi_{\mu j} - \nonumber \\
  && - \frac12 \varepsilon^{ij} \bar{\chi}_i \gamma^\mu \gamma^\nu
\left[ \delta^k_j \partial_\nu \tilde{\varphi} + e^{\tilde{\varphi}}
\partial_\nu (\pi)^k_j \right] \Psi_{\mu k} + \nonumber \\
  && + \frac{i}{4\sqrt{2}} e^{-\sqrt{2}\varphi} \left\{
\varepsilon^{\mu\nu\alpha\beta} \bar{\Psi}^i_\mu \gamma_\nu \Psi_{\beta
i} + 3 \bar{\Omega}^i \gamma^\alpha \gamma_5 \Omega_i - \bar{\chi}^i
\gamma^\alpha \gamma_5 \chi_i \right\} \partial_\alpha \pi - \nonumber  \\
  && - \frac{i}4 e^{\tilde{\varphi}} \left\{
\varepsilon^{\mu\nu\alpha\beta} \bar{\Psi}^i_\mu \gamma_5 \gamma_\nu
\partial_\alpha \pi^j_i \Psi_{\beta j} + \bar{\Omega}^i \gamma^\mu
\partial_\mu \pi^j_i \Omega_j - \bar{\chi}^i \gamma^\mu \partial_\mu
\pi^j_i \chi_j \right\}
\end{eqnarray}
This Lagrangian is invariant under the following local
supertransformations (up to the terms bilinear in fermions)
\begin{eqnarray}
 \delta e_{\mu r} &=& i(\bar{\Psi}^i_\mu \gamma_r \eta_i) \nonumber \\
 \delta \Psi_{\mu i} &=& 2D_{\mu} \eta_i + \frac{i}4
e^{\varphi/\sqrt{2}} \varepsilon_{ij} \sigma (A - \gamma_5 B) \gamma_\mu
\eta^j + \nonumber \\
  && + \frac1{\sqrt{2}} e^{-\sqrt{2}\varphi} \gamma_5 \partial_\mu \pi
\eta_i - e^{\tilde{\varphi}} \partial_\mu \pi^j_i \eta_j \nonumber \\
 \delta A_{\mu} &=& e^{-\varphi/\sqrt{2}} \left[ - \varepsilon^{ij}
(\bar{\Psi}_{\mu i} \eta_j) + \frac{i}{\sqrt{2}} (\bar{\Omega}^i
\gamma_\mu \eta _i) \right] \nonumber \\
 \delta B_{\mu} &=& e^{-\varphi/\sqrt{2}} \left[ - \varepsilon^{ij}
(\bar{\Psi}_{\mu i} \gamma_5 \eta_j) + \frac{i}{\sqrt{2}}
(\bar{\Omega}^i \gamma_\mu \gamma_5 \eta _i) \right] \\
 \delta \Omega_i &=& - \frac1{2\sqrt{2}} e^{\varphi/\sqrt{2}} \sigma (A
+ \gamma_5 B) \eta _i -i \varepsilon_{ij} \gamma_\mu \left[ \partial_\mu
\varphi + e^{- \sqrt{2}\varphi} \gamma_5 \partial_\mu \pi \right] \eta^j
\nonumber \\
 \delta \varphi  &=& \varepsilon^{ij} (\bar{\Omega}_i \eta_j)\qquad
\delta \pi = e^{\sqrt{2}\varphi} \varepsilon^{ij}(\bar{\Omega}_i
\gamma_5 \eta_j)\nonumber \\
 \delta \chi_i &=& - \varepsilon_{ij} \gamma_\mu \left[ \delta^k_j
\partial_\mu \tilde{\varphi} + e^{\tilde{\varphi}} \partial_\mu \pi^k_j
\right] \eta _k \nonumber \\
 \delta  \tilde{\varphi} &=& \varepsilon^{ij} (\bar{\chi}_i \eta_j)
\qquad \delta \pi^a = e^{- \tilde{\varphi }} \varepsilon^{ij}
(\bar{\chi}_i(\tau^a)^k_j \eta_k) \nonumber
\end{eqnarray}

  At a standard "minimal" interaction of the vector multiplet with the
$N=2$ supergravity the equations of motions are known \cite{r7} to be
invariant under the global (non-linearly realized) $SU(1,1)$ group,
which includes dual transformations of the vector fields. Here the
Lagrangian is invariant under the non-compact $O(1,1)$ subgroup. As is
seen from (5) and (6) the version considered is obtained from a standard
one through the dual transformation, and, in this, Lagrangian (5) is
invariant under the compact subgroup $O(2)$.

  Peculiar requirements for the linear multiplet have brought us to the
fact that its interaction with the $N=2$ supergravity differs, in
principle, from the "standard" one, which is obtained in the framework
of the conformal tensor calculus (see, eg. \cite{r8}). Right this
circumstance allows one to obtain a partial symmetry breaking without a
cosmological term in our model, while it turns out to be impossible for
a wide class of models based on the conformal tensor calculus \cite{r9}.

\section{}

  As it has been noted in the Introduction, supergravity interaction
with matter is naturally constructed in terms of massless
supermultiplets. In extended supergravities the introduction of mass
terms, as well as of Yukawa interactions and scalar field potentials is
unambigously connected with switching of gauge interaction. In such
theories the replacement of the derivatives by covariant ones with
respect to some gauge group $G$ breaks the invariance of the Lagrangian
under supertransformations. This invariance may be restorted if the terms
of the Yukawa type interaction and scalar field potential are added to
the Lagrangian, while the corresponding terms are added to the fermion
transformation laws. In this, vacuum expectation values of the scalar
fields corresponding to the minimum of the potential define the mass
spectrum, the presence or absence of symmetry breaking, etc. In
particular, this is the way one manages to reproduce the interaction of
massive vector and scalar supermultiplets with the $N=2$ supergravity
\cite{r10}.

  Attempts to obtain spontaneously broken supersymmetry using
non-Abelian gauge group, result in arising cosmological term. Its
absence in our model is connected with use of Abelian non-compact gauge
transformations. Since the fields $\pi^a$ enter only in the form
$\partial_\mu \pi^a$ the Lagrangian (5) is invariant under the global
translations $\pi^a \rightarrow \pi^a + \Lambda^a$. This allows one
using two of these fields (eg. $\pi^1$ and $\pi^2$) to introduce masses
to two Abelian vector fields $A_\mu$ and $B_\mu$. As is seen from (1),
the case of equal masses corresponds to partial breaking of
supersymmetry. However one can treat a general case when $m_1 \ne m_2$,
making the following replacements in the Lagrangian (5) and
supertransformations (6):
\begin{equation}
 \partial_\mu \pi^j_i \to  \partial_\mu \pi^j_i - \left(
\begin{array}{cc}  O & - m_1 A_\mu + m_2 \gamma_5 B_\mu \\ m_1 A_\mu +
m_2 \gamma_5 B_\mu & O \end{array} \right)^j_i
\end{equation}
As usually such a replacement violates the invariance of the Lagrangian
under the supertransformations, therefore the Lagrangian should be
supplemented with
\begin{eqnarray}
 \Delta L &=& e^{(\tilde{\varphi} - \varphi / \sqrt{2})} \{ \frac14
\bar{\Psi}_{\mu i} \sigma^{\mu\nu} \left( \begin{array}{cc} m_1 & -m_2 O
\\ O & m_1  + m_2 \end{array} \right)^{ij} \Psi_{\nu j} + \nonumber \\
  && + \frac{i}{2\sqrt{2}} \Psi^i_\mu \gamma^\mu \left( \begin{array}{cc}
 O & m_1 - m_2 \\ - m_1 - m_2 & O \end{array} \right)^j_i \Omega_j -
\frac{i}2 \Psi^i_\mu \gamma^\mu \left( \begin{array}{cc}  O & m_1
- m_2 \\ - m_1 - m_2 & O \end{array} \right)^j_i \chi_j - \nonumber \\
  && - \frac1{\sqrt{2}} \bar{\Omega}_i \left( \begin{array}{cc}  m_1
+ m_2 & O \\ O & m_1 - m_2 \end{array} \right)^{ij} \chi_j  +
\frac14 \bar{\chi}_i \left( \begin{array}{cc} m_1  + m_2 & O \\ O & m_1
- m_2 \end{array} \right)^{ij} \chi_j\}
\end{eqnarray}
and supertransformations (6) with
\begin{eqnarray}
 \delta '\Psi_{\mu i} &=& - \frac{i}2 e^{(\tilde{\varphi} -
\varphi/\sqrt{2})} \gamma_\mu \left( \begin{array}{cc} m_1 - m_2 & O \\
 O & m_1 + m_2 \end{array} \right)^{ij} \eta_j \nonumber \\
 \delta '\Omega_i &=& \frac1{\sqrt{2}} e^{(\tilde{\varphi} -
\varphi/\sqrt{2})} \left( \begin{array}{cc} O & - m_1 - m_2 \\ m_1 - m_2
 & O \end{array} \right)^j_i \eta_j \\
 \delta '\chi_i &=& - e^{(\tilde{\varphi} - \varphi/\sqrt{2})} \left(
\begin{array}{cc} O & - m_1 - m_2 \\ m_1 - m_2 & O \end{array}
\right)^j_i \eta _j\nonumber
\end{eqnarray}
Formulae (8) and (9) are unambigously determined by the requrements of
the invariance of the total Lagrangian (5)+(8) respect to
supertransformations (6)+(9).

  From (8) it is seen that in the general case $m_1 \ne m_2$
supersymmetry is completely broken, in this the gravitini masses are
connected with the vector fields masses through the relations
\begin{equation}
 \mu_1 = \frac{m_1 - m_2}2 \qquad \mu_2 = \frac{m_1 + m_2}2
\end{equation}
Besides two goldstone spinors there two spinor fields with the same
masses $\mu_1$ and $\mu_2$ in the model. It is interesting to note that
in this case a well known relation is fulfilled
\begin{equation}
 \Sigma (-1)^{2s} (2s + 1) m^2_s = O
\end{equation}
where the sum is taken over all massive fields with the spin $s$. In
particular case $m_1 = m_2$, one gravitino remaines massless, and the
corresponding supertransformation --- unbroken.

  From (8) it is also clear that the scalar field potential in the given
minimal model does not arise at all, and as a consequence the
cosmological term is automatically equals to zero.

  Thus for the $N=2$ supergravity we managed to construct a system,
where there takes place a spontaneous supersymmetry breaking with two
arbitrary scales, one of which may be equal to zero, that corresponds to
a partial super-Higgs effect. It would be interesting to consider the
interaction of this system (as a "hidden" sector) with matter, i.e. with
vector ans scalar multiplets. This question will be dealt with in our
future publications.

\end{document}